\def\year{2022}\relax
\documentclass[letterpaper]{article} 
\usepackage{aaai22}  
\usepackage{times}  
\usepackage{helvet}  
\usepackage{courier}  
\usepackage[hyphens]{url}  
\usepackage{graphicx} 
\usepackage{natbib}  
\usepackage{caption} 
\DeclareCaptionStyle{ruled}{labelfont=normalfont,labelsep=colon,strut=off} 
\usepackage{cite}
\usepackage{amsmath,amssymb,amsfonts}
\usepackage{algorithmic}
\usepackage{graphicx}
\usepackage{textcomp}
\usepackage{xcolor}
\usepackage{booktabs}
\usepackage{tabularx}
\def\BibTeX{{\rm B\kern-.05em{\sc i\kern-.025em b}\kern-.08em
    T\kern-.1667em\lower.7ex\hbox{E}\kern-.125emX}}
\usepackage{tikz}
\usetikzlibrary{shapes.geometric, arrows}
\tikzstyle{startstop} = [rectangle, rounded corners, minimum width=3cm, minimum height=1cm,text centered, draw=black, fill=red!30]
\tikzstyle{io} = [trapezium, trapezium left angle=70, trapezium right angle=110, minimum width=3cm, minimum height=1cm, text centered, draw=black, fill=blue!30]
\tikzstyle{process} = [rectangle, minimum width=3cm, minimum height=1cm, text centered, draw=black, fill=orange!30]
\tikzstyle{decision} = [diamond, minimum width=3cm, minimum height=2cm, text centered, draw=black, fill=green!30]
\tikzstyle{arrow} = [thick,->,>=stealth]

\usepackage{todonotes}
\usepackage{csquotes}
\usepackage{siunitx}
\usepackage{enumitem}
\setitemize{leftmargin=*}
\usepackage{caption}
\usepackage{endnotes}
\usepackage{amsmath}
\usepackage{amssymb}
\usepackage[ruled, vlined, linesnumbered]{algorithm2e}
\usepackage{algorithmic}
\usepackage{listings}
\usepackage{url}

\newcommand\hide[1]{}
\newboolean{show_notes}
\setboolean{show_notes}{true}
\newcommand\note[1]{\ifthenelse{\boolean{show_notes}}{\textcolor{red}{\textbf{Note: }#1}}{\hide{#1}}}

\title{Evolutionary Hierarchical Harvest Schedule Optimization\\for Food Waste Prevention}
\author {
    Maurice Günder\textsuperscript{\rm 1,2,*},
    Nico Piatkowski\textsuperscript{\rm 2},
    Laura von Rueden\textsuperscript{\rm 2},
    Rafet Sifa\textsuperscript{\rm 2},
    Christian Bauckhage\textsuperscript{\rm 1,2}\\
}
\affiliations {
    \textsuperscript{\rm 1} Institute for Computer Science III, University of Bonn, Friedrich-Hirzebruch-Allee 5, 53115 Bonn, Germany \\
    \textsuperscript{\rm 2} Fraunhofer Institute for Intelligent Analysis and Information Systems IAIS, Schloss Birlinghoven, 53757 Sankt Augustin, Germany \\
    \{maurice.guender; nico.piatkowski; laura.von.rueden; rafet.sifa; christian.bauckhage\}@iais.fraunhofer.de \\
    \textsuperscript{\rm *} Corresponding author
}

\begin{document}

\maketitle

\begin{abstract}
In order to avoid disadvantages of monocropping for soil and environment, it is advisable to practice intercropping of various plant species whenever possible. However, intercropping is challenging as it requires a balanced planting schedule due to individual cultivation time frames. Maintaining a continuous harvest reduces logistical costs and related greenhouse gas emissions, and contributes to food waste prevention. In this work, we address these issues and propose an optimization method for a full harvest season of large crop ensembles that complies with given constraints. By using an approach based on an evolutionary algorithm combined with a novel hierarchical loss function and adaptive mutation rate, we transfer the multi-objective into a pseudo-single-objective optimization problem and obtain faster convergence and better solutions than for conventional approaches.
\end{abstract}

\section{Introduction}\label{chap:intro}

In a recent study, the Food and Agriculture Organization of the United Nations stated that almost \num{1.4} billion hectares or \SI{30}{\percent} of the world's agricultural land are used to produce food that later goes to waste. They also stated that global food loss causes \num{3.3} gigatons of carbon dioxide emissions or \SI{7}{\percent} of the world's total greenhouse gas emissions~\cite{fao_foodwaste_report, fao_sofa_2019}. Efficient use of agricultural and logistical capacities is therefore crucial for sustainable global food supply.

\begin{figure}[t]
	\centering
	\includegraphics[width=0.75\columnwidth]{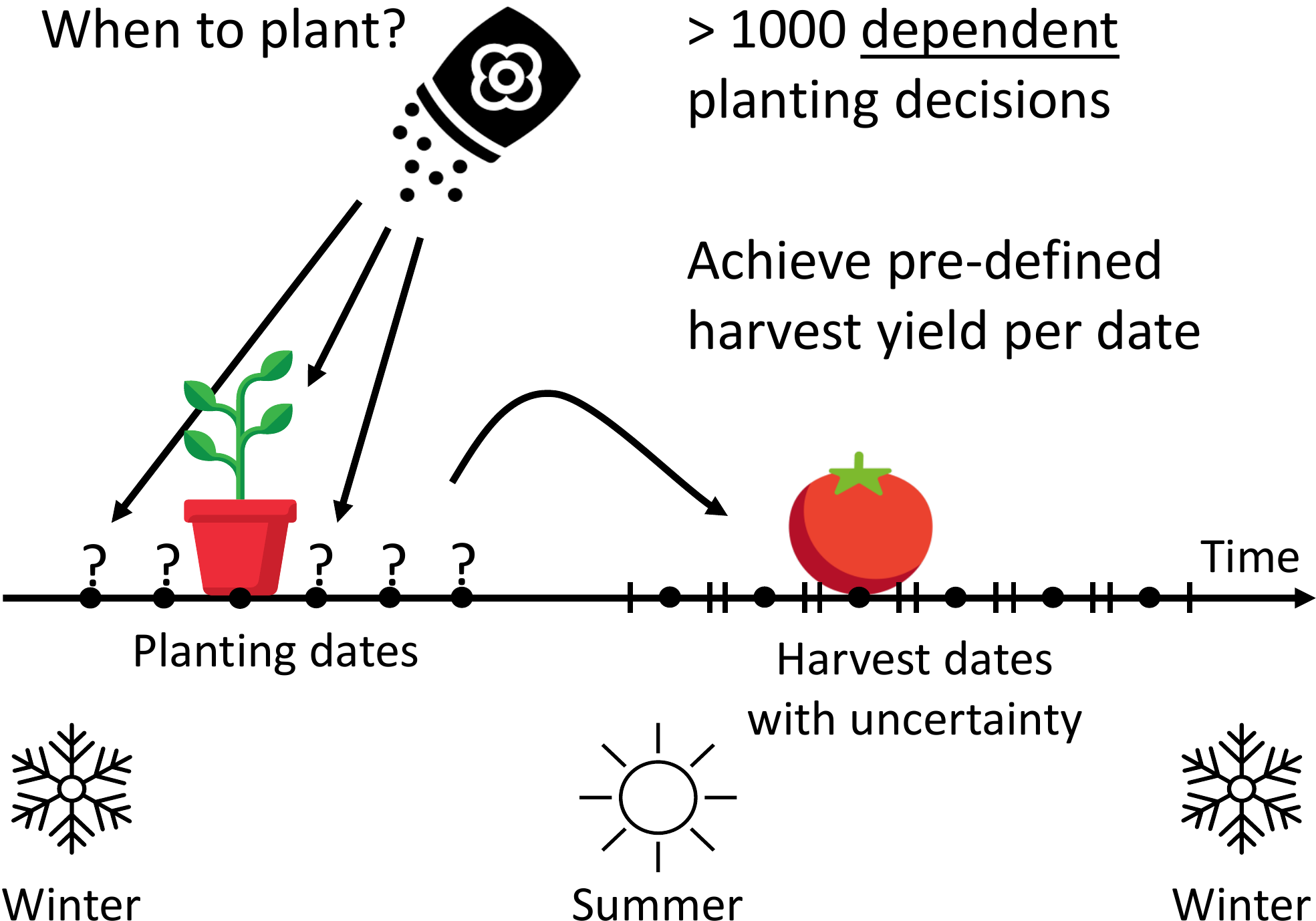}
	\caption{Problem setting: for given planting dates, corresponding harvest dates have to be predicted. Predictions are based on uncertain Growing Degree Unit (GDU) forecasts. The challenge is to schedule planting dates for more than \num{1000} individual crops and desired yields simultaneously, resulting in a continuous harvest yield distribution.}
	\label{fig:problem_sketch}
\end{figure}

A particular application of artificial intelligence in agriculture is the optimization of harvest schedules or agri-food supply chains~\cite{survey_harvest_optimization}. Even in other food-related disciplines, harvest schedules are of major importance, e.g. in aquaculture~\cite{harvest_scheduling_aquaculture}. In this work, we will focus on optimizing a planting schedule with regard to a sustainable yield sketched in Figure~\ref{fig:problem_sketch}. In particular, our goal is to find planting dates for many crop species such that the resulting yield is efficiently distributed. In this way, we facilitate prevention of food waste and surplus greenhouse gas emissions.

\begin{figure}[t]
	\centering
	\includegraphics[width=0.75\columnwidth]{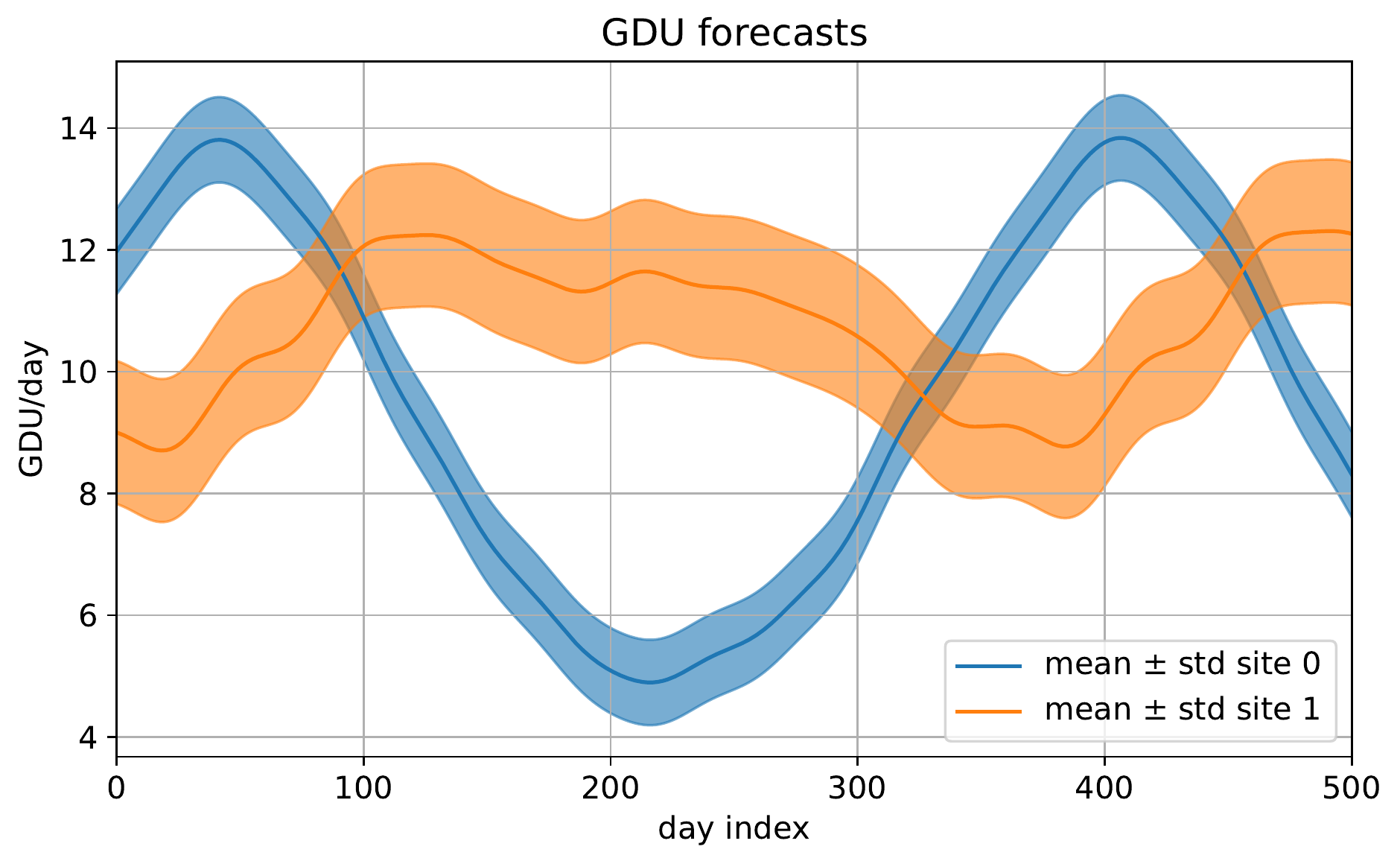}
	\caption{GDU forecast mean and $\pm 1\sigma$ standard deviation for site 0 (blue) and site 1 (orange).}
	\label{fig:gdu_forecast}
\end{figure}

Scheduling problems are difficult to solve by involving numerous soft- and hard constraints and requiring to factor in environmental conditions. Common approaches thus consider stochastic models and commercial solvers which, however, are essentially \enquote{black box} systems whose decisions are opaque and difficult to comprehend. Alternatively, scheduling problems are commonly cast as multi- or many-objective optimization~\cite{moo_book} problems. Besides many other optimization strategies like in \cite{hs-opt_winegrapes} and \cite{hs-opt_multiobjective2}, these problems are often optimized by evolutionary algorithms (EAs) like NSGA-II~\cite{nsga2} or MOEA/D~\cite{moead} as in \cite{hs-opt_multiobjective1}, or in combination with other models \cite{hs-opt_bilevel_ea}.

In the work at hand, we consider a basic evolution strategy (ES) often referred to as $(1+1)$-ES. It is simple to implement and can be generalized towards $(\mu+\lambda)$-ES which allow for distributed computation.

Our contributions can be summarized as follows:
\begin{itemize}
    \item we propose a generic formalization of the harvest schedule optimization problem, which is agnostic with respect to time scales, crop types, and desired yield
    \item a novel hierarchical loss function is devised that converts naive multi-objective problems into a single-objective problem, enabling faster optimization
    \item uniform escape from local optima is facilitated via a new dynamic oscillating mutation rate
    \item we provide an experimental evaluation on realistic data
\end{itemize}

This work provides a summary of results for the above points. A detailed discussion of methods and the full set of results will be available in an extended version of this manuscript.

\section{Background}\label{chap:notation_background}

\begin{figure*}[!htb]
	\centering
	    \includegraphics[width=0.3\textwidth]{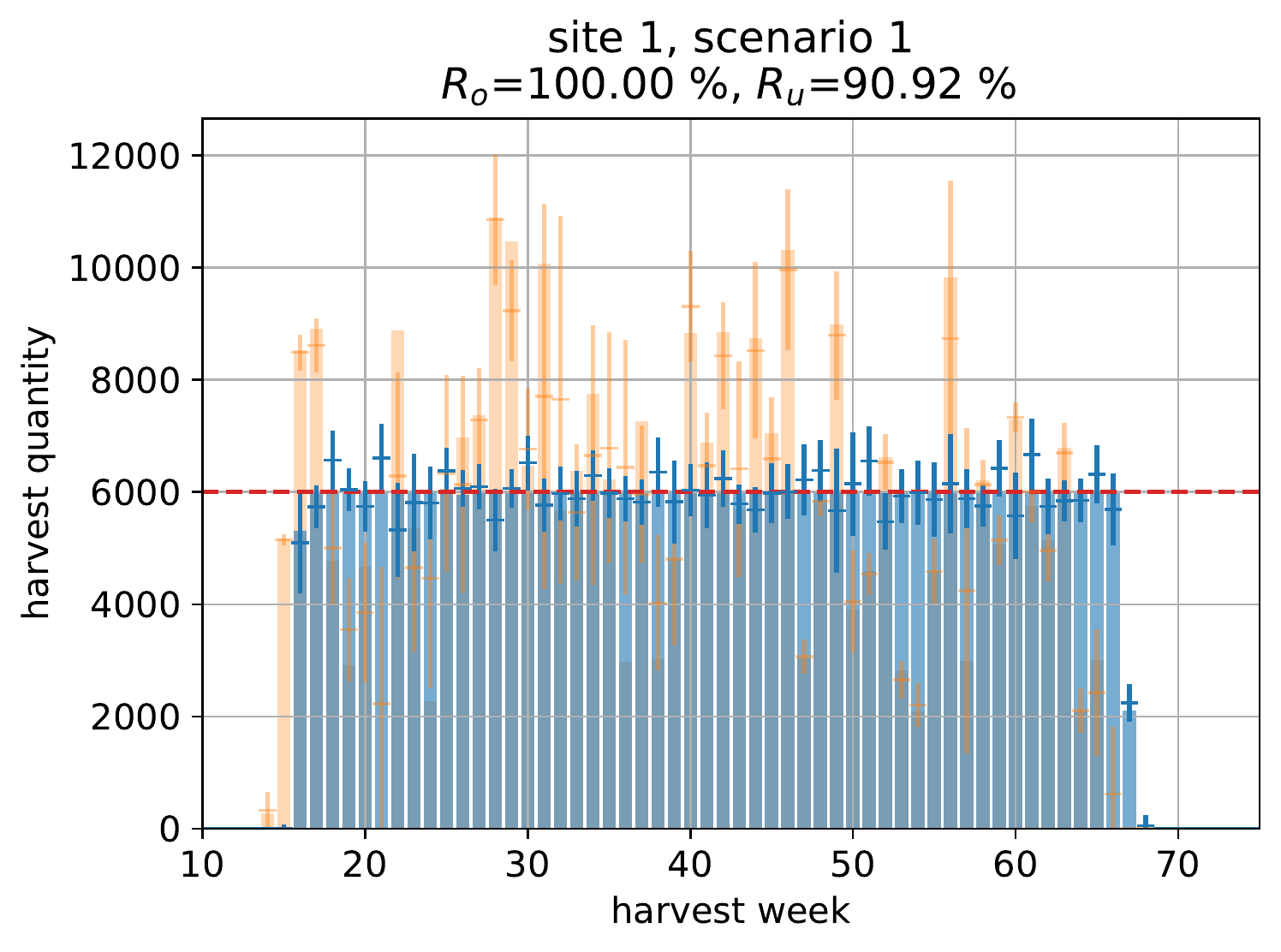}
	\hfill
		\includegraphics[width=0.3\textwidth]{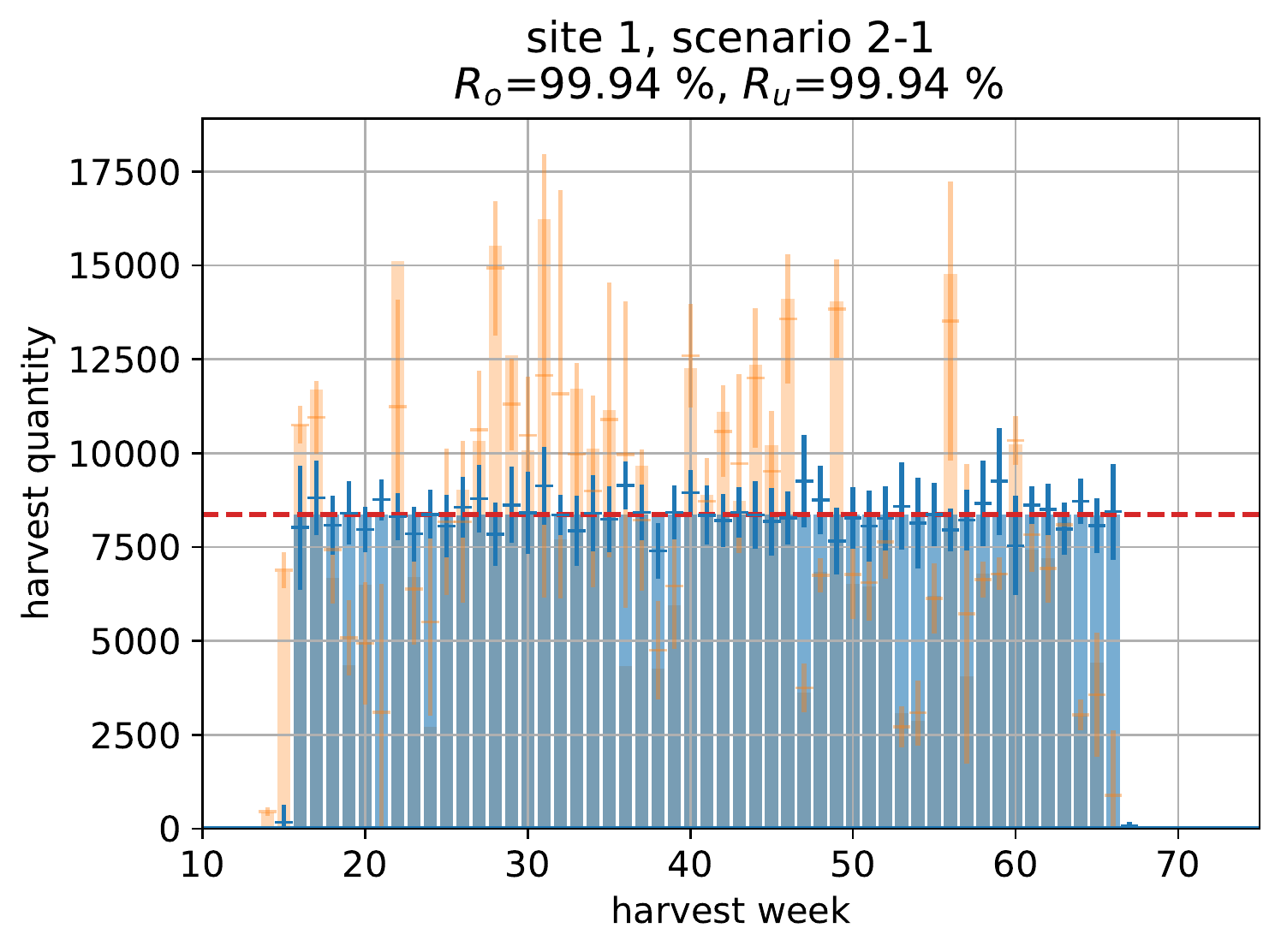}
    \hfill
    	\includegraphics[width=0.3\textwidth]{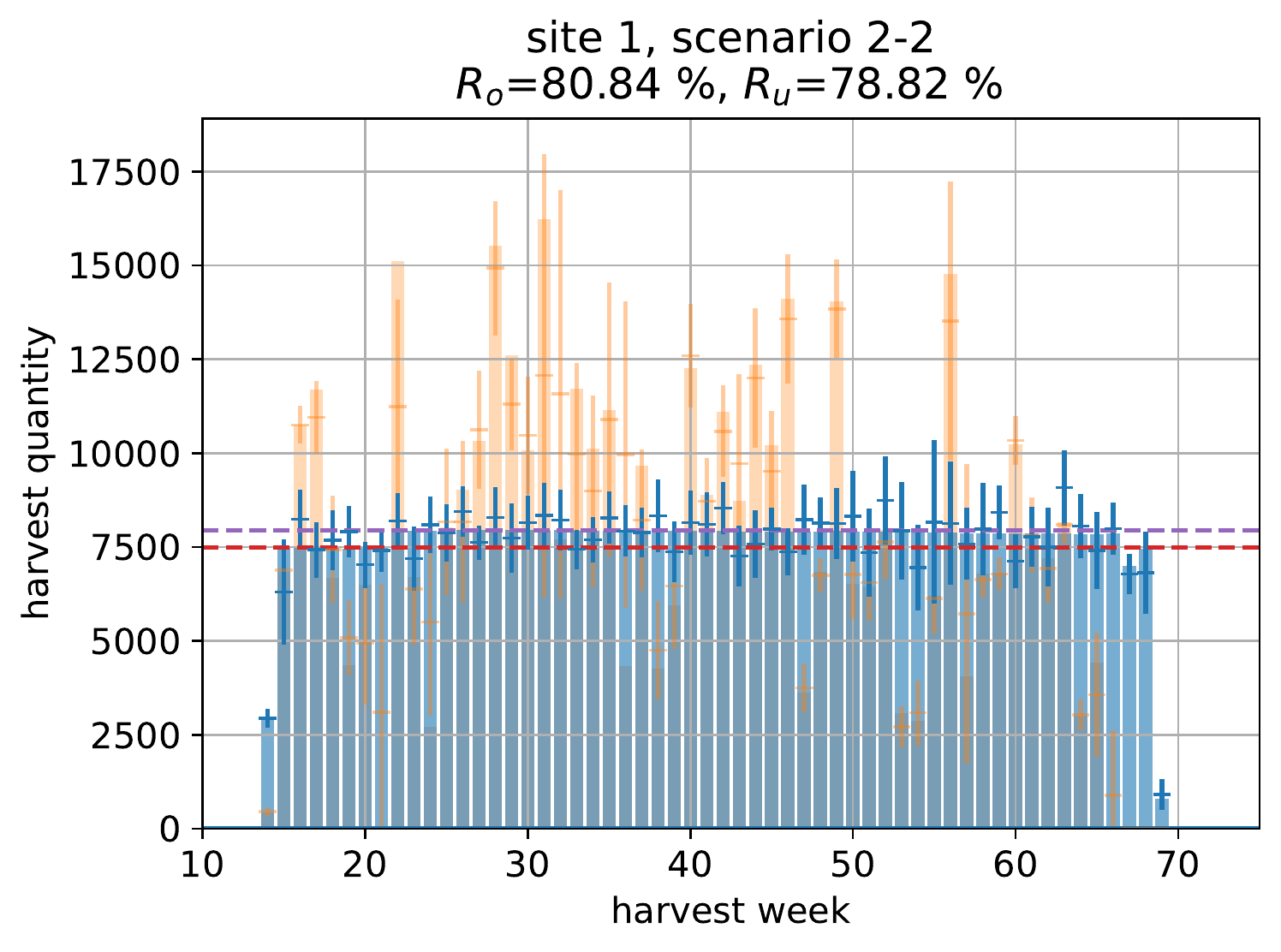}
	\caption{Weekly harvest quantity for all 3 scenarios on site 1. The bars show the harvest per week for an unoptimized schedule given in the dataset (orange) and our results (blue). The dashed lines mark target (red) and maximally needed capacity (purple). For uncertainty estimation (error bars), \num{100} harvest matrices are resampled from the GDU forecast. The over- and undershoot reduction ratios $R_o$ and $R_u$ refer to the unoptimized schedule. It is visible that our optimized schedule largely prevents over- and undershoot.}
	\label{fig:harvest_results}
\end{figure*}

The first building block of our method is Gaussian process regression (GPR) for time series prediction \cite{gp_rasmussen}. GPR is a kernel-based non-parametric method to inter- and extrapolate data points. It allows us to include available background knowledge about agricultural time series, e.g. the existence of annual periodicity, other seasonal effects, and trends, by choosing appropriate kernel functions. 

The second building block are evolutionary algorithms (EAs) which are a class of meta-heuristic optimization algorithms inspired by biological mechanisms \cite{ea_book1, ea_book2}. The common procedure is to generate one or more possible solutions in the form of state vectors, called \textit{population}. During a \textit{generation}, the populations, or \textit{parents}, are \textit{mutated}, resulting in \textit{children}. An \textit{environment} thereafter has to select the best population(s) which themselves are initial states for further mutations in the upcoming generation \cite{ga_parameter_review}. This procedure within a generation cycle is referred to as \textit{evolution strategy} (ES). We focus on $(\mu +\lambda)$-ES \cite{es_overview}. $\mu$ is the number of parents that undergo mutations resulting in $\lambda$ children. The \enquote{$+$} sign indicates that, when selecting the $\mu$ \enquote{fittest} mutants, parents \textit{and} children are taken into account. Since EAs are meta-heuristic optimization methods, they typically yield only approximately optimal solutions to a problem. The major advantage, however, is that they can handle discrete high-dimensional and multi-objective inputs and that their objective- or \textit{fitness} function does not have to be (continuously) differentiable as in the case of, say, gradient descent based methods.

\section{Modeling the Problem}\label{chap:methodology}
Next, we focus on our harvest scheduling problem by discussing constraints, the mathematical problem formalization, and adaptations to the optimization algorithm, i.e. oscillating mutation rate and the hierarchical loss.

\subsection{Constraints}\label{sec:constraints}
Seeding and growing of crop are fraught with several constraints and uncertainties. Two constraints are the windows or periods of possible planting and their planting as a whole at one point in time for each individual crop. Furthermore, we introduce the two variables, i.e. desired yield for each species and the \enquote{performance} of the field to accumulate crop growth. In phenology, the latter is often measured by \textit{Growing Degree Units}. This is the accumulated time-dependent temperature $T(t)$ above a certain base temperature $T_\text{base}$ during the interval of interest. Hence, $GDU = \int (T(t)-T_\text{base}) dt$. Commonly, the interval is a day, which is why it is also referred to as Growing Degree Days (GDD)~\cite{gdu_paper}. Seasonal (and geographical) variations in GDU accumulation have direct impact on the predicted harvest time. Thus, the harvest time is a probability density over time points, although further simplifications like mean harvest time and standard deviations or a simple time interval are conceivable as well. EAs work on discrete domains, so we will consider discrete time points and intervals. They do not necessarily have to be equidistant, but one usually considers days or weeks as units.

\subsection{Formalization}
Next, we introduce quantities that impact our problem. Let $\mathcal{D}=\{1,2,\dots,d_\text{max}\}\subset\mathbb{N}^+$ be the possible planting and harvesting dates. The dates themselves are encoded with an index $d\in\mathcal{D}$ as well as the earliest and latest planting dates for each species $s\in\mathcal{S}$, thus $d_\text{early}(s), d_\text{late}(s)\in\mathcal{D}$. We further introduce $\mathcal{S}$ as the set of species. Moreover, we have the required accumulated amount of GDU for each species to be harvested, i.e.~$g_\text{harvest}(s)$. Thus, we calculate the harvest date by $ d_\text{harvest}(s, d_\text{plant}) =$
\begin{align}\label{eq:havest_date}
    \underset{d}{\min{}} \left\{ d \geq d_\text{plant}\;\middle|\;\sum_{x=d_\text{plant}}^{d} g_\text{acc}(x) \geq g_\text{harvest}(s)\right\}\,,
\end{align}
with GDU accumulation function $g_\text{acc}(x)$.

Given all this information, we define a harvest matrix $\mathbf{H}\in\mathcal{D}^{|\mathcal{S}|\times |\mathcal{D}|}$ that contains the harvest date for each species (as rows) and each planting date (as columns). We assign the value $-1$ impossible planting dates. Thus, $\mathbf{H} = \{H(i,j)\}$ where $H(i,j) =$
\begin{align}\label{eq:harvest_matrix}
    \begin{cases}
    d_\text{harvest}(s=i, d_\text{plant}=j) & \text{if} \; j\in[d_\text{early}(i),d_\text{late}(i)]\\
    -1 & \text{otherwise}
    \end{cases}\,,
\end{align}
with harvest dates from Equation~\eqref{eq:havest_date}.

The precalculation of the harvest matrix reduces the schedule calculation time. With $\vec{q}=(q(s))\in(\mathbb{R}_0^+)^{|\mathcal{S}|}$ as the vector of desired harvest quantity for each crop species, we can write the date harvest $\vec{h}(\vec{d}_\text{plant})\in(\mathbb{R}_0^+)^{|\mathcal{D}|}$ as
\begin{align}\label{eq:harvest_schedule_calc}
\vec{h}(\vec{d}_\text{plant}) =  \sum_{s\in\mathcal{S}} \Big\{ \mathbf{I}_{d_\text{max}}(\mathbf{H}(s,d_\text{plant}(s)),d)q(s) \Big\}_{d\in\mathcal{D}}\,,
\end{align}
where $\mathbf{I}_{d_\text{max}}$ is the identity matrix with $d_\text{max}$ dimensions mapping dates to themselves again. For instance, a date-to-week mapping matrix yields the weekly harvest.

\subsection{Designing the Optimization Algorithm}
In this work, we consider the $(1+1)$-ES. This turns out to be a good choice in our experiments, and there is no optimal choice due to the \textit{No-free-Lunch-Theorem for Search} \cite{nofreelunch_theorem}. In the case that parent and child have the same loss value, we prefer the child to the parent, increasing the probability to overcome \textit{plateaus} in the loss landscape.

\subsubsection{Oscillating Mutation Rate.}
We draw the initial planting dates from a uniform distribution $\mathcal{U}$ by $d_\text{plant}(s)~\sim~\mathcal{U}[d_\text{early}(s), d_\text{late}(s)]$ for each species $s\in\mathcal{S}$ within its individual time window. Mutation is done by redrawing few elements. Further, we will use an initial mutation rate of $\frac{1}{|\mathcal{S}|}$, so we expect one planting to change by mutation. After this initial step, we exploit the advantages of an adaptive mutation rate~\cite{adaptive_mutation_1,adaptive_mutation_2} by introducing a counter $j=0$ that is increased by $1$ if the mutant is worse than the parent. This is done up to a maximum ratio $\rho_\text{max}$ of elements. If the mutant is better than the parent, the counter is reset to $0$. Moreover, there is the mentioned \enquote{plateau case} where the mutant is as good as the parent. In this case, the counter keeps its current value. Hence, we formulate a mutation rate that is oscillating between $\frac{1}{|\mathcal{S}|}$ and $\rho_\text{max}$ with frequency $\omega$.
\begin{align}\label{eq:mutation_rate}
\rho(j) = \frac{1}{|\mathcal{S}|}\bigl( 1+(\rho_\text{max}|\mathcal{S}|-1) \sin^2(\omega j) \bigr)\,,
\end{align}
which enables the algorithm to try a wide range of mutations and escape from local optima without turning into a random walk. This principle is connected with the concept of \enquote{temperature} in simulated annealing \cite{sa_paper} but with advantage to be adaptive to the landscape in vicinity of the current best solution.

\subsubsection{Hierarchical Loss Function.}\label{sec:loss}
First, we define the loss vector $\vec{L}(\vec{h}) = (L_+, L_-) = \left( \sum_{h\in\vec{h}_+} l_C(h), \sum_{h\in\vec{h}_-} l_C(h) \right)$ with
\begin{align}
l_C(h) &=
\begin{cases}
\frac{h}{C}\left(1-\frac{h}{C}\right)\, & \frac{h}{C} < 1\\
\exp{\left(\frac{h}{C}\right)}-\exp(1)\, & \frac{h}{C} \geq 1
\end{cases}\,,\label{eq:loss_function}
\end{align}
which has to be minimized in the following. $\vec{h}_+$ only contains harvest quantities above and equal to the capacity limit, and $\vec{h}_-$ only contains those below. We discriminate between two cases with different optimization goals. For $h(d) < C_\text{target}$, we want either full capacity harvest or no harvest at all, whereas for $h(d) \geq C_\text{target}$, we want to minimize overshoot. The total harvest is constant, meaning that decrease of harvest at one date simultaneously increases harvest at another date by the same quantity. Our loss promotes high-harvest dates getting \enquote{full} in terms of the target capacity, while simultaneously eliminating low-harvest dates. Thus, in case of $h < C_\text{target}$, two harvest quantities \textit{diverge}, provoked by a concave loss function. For $h \geq C_\text{target}$, the involved dates' quantities \textit{converge}, provoked by a \textit{convex} function.

Although this might sound like a multi-objective approach, we still formulate a single-objective problem by preferring the loss in the region above capacity limit ($L_+$) over the region below it ($L_-$). Thus, every net change from over into under capacity is promoted. This is what we call \textit{hierarchical loss} where hierarchy decreases with increasing element index in $\vec{L}$. The flexibility to include further objectives or knowledge at this stage is obvious since one may add loss categories.

\section{Application and Results}

In the considered dataset, two harvesting sites 0 and 1 are available. We will perform the optimization along two scenarios S1 and S2, targeting different objectives. In S1, the sites have weekly harvest capacity limits, whereas in S2, the capacity is not given. Here, a reasonable capacity should be found during optimization. The GDU accumulation for harvesting season 2020/21 is forecasted by using a historical dataset of \num{11} years for both planting sites. The accumulation curves are plotted in Figure~\ref{fig:gdu_forecast}.

The GDU accumulation forecast comes along with uncertainties. Fortunately, GPR models have the property of not just extrapolating unseen regions, but also giving the confidence of their predictions. The actual predicted value is a mean value with a Gaussian standard deviation. We can exploit this to propagate uncertainties and estimate the sensitivity of our optimization to different GDU accumulation scenarios, i.e. weather conditions, by resample many harvest matrices by the bootstrap technique~\cite{bootstrap} and evaluate our result for all. We further substitute the identity matrix in Equation~\eqref{eq:harvest_schedule_calc} for date-to-date mapping with a day-to-week mapping binary week matrix $\mathbf{W}\in\{0;1\}^{|\mathcal{D}|\times|\mathcal{W}|}$.

We further compared our method with two established EAs, namely NSGA-II and MOEA/D, by using two multi-objective optimization strategies each. For a short-time run of $10^6$ generations, all those variants are outperformed by our method. Additionally, we formulated the problem as a mixed-integer linear program (MILP) and solved it with CPLEX\textregistered{}. Here, we observed as well that our method leads to significantly better results than the exact one of the solver. For the sake of brevity, a more detailed elaboration and plots can be found in the Appendix.

\subsection{Harvest Schedule Optimization}\label{sec:optimization}
For the optimization scenarios, our proposed EA applies the adaptive mutation rate shown in Equation~\eqref{eq:mutation_rate}. We use $\rho_\text{max} = \SI{1}{\percent}$ and $\omega = \num{5e-4}$. The number of species $|\mathcal{S}|$ is \num{1375} for site 0 and \num{1194} for site 1. 

Our hierarchical loss needs the capacity limit for the separation between the two regions. For S1, the target capacity is given. We split S2 into two slightly different approaches, S2-1 and S2-2, since \enquote{lowest capacity needed} is always a trade-off with the number of total harvest weeks. The optimization is exactly the same as for S1 but with different inference of the capacity limit. In S2-1, we link the target capacity limit to the number of harvest weeks the current solution needs. For S2-2, we determine the maximum number of possible harvest weeks. Thus, we expect S2-1 to need a higher capacity in less weeks whereas S2-2 needs a lower capacity in more weeks if not all possible weeks are harvest weeks after all.

In Figure~\ref{fig:harvest_results}, the weekly harvest quantity is plotted against the harvest weeks for site 1. The light yellow bars are the weekly harvest quantity if the plants are planted by the original planting schedule given in the data. Blue bars represent our optimized result after $10^9$ generations.

\section{Discussion}

In the following, we illuminate our optimization results from a technical as well as a social impact point of view.

\subsection{Optimization Process}
\begin{figure}
	\centering
	\includegraphics[width=0.85\columnwidth]{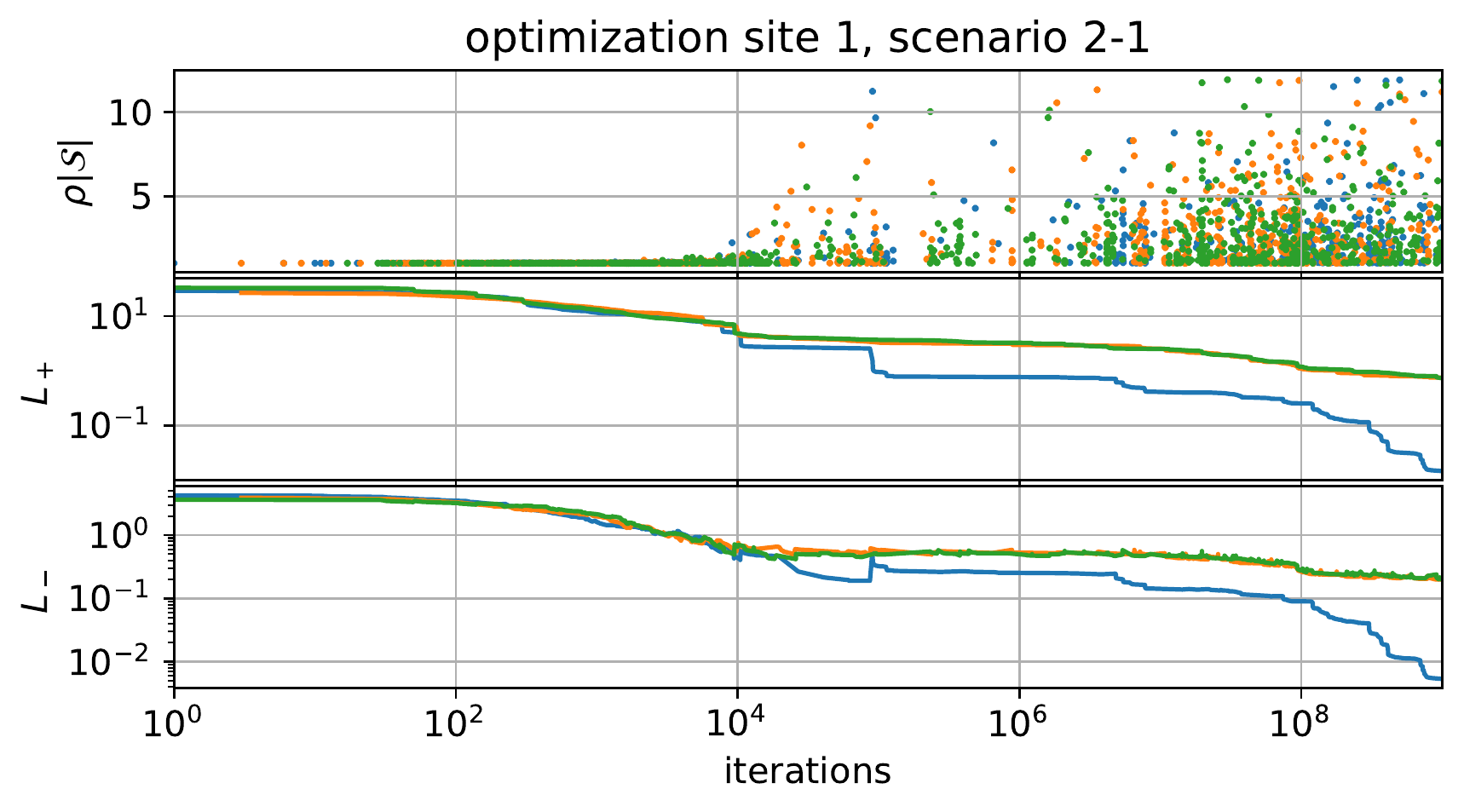}
	\caption{Optimization process with 3 independent runs. The two lower plots show the loss vector components $L_+$ and $L_-$ against iterations. The upper plot shows the mutation rate for each successful improvement as a dot.}
	\label{fig:loss_hist}
\end{figure}

All scenarios show that the main optimization process needs orders of $10^6$generations to saturate at a moderately well optimized state. The optimization process plot in Figure~\ref{fig:loss_hist} shows that behavior exemplarily for two scenarios and three independent runs each. Beyond $10^7$ generations, fine-tuning occurs, e.g., if a constellation is found that a complete harvest week can be canceled for. The optimization process directly shows the characteristics and the strength of our method. Firstly, the hierarchical character of the loss is observable by a monotonically decreasing $L_+$ and an $L_-$ showing jumps to higher values. Secondly, fine-tuning preferably happens with higher mutation rates than the intermediate saturation. Using the adaptive mutation rate prevents the EA from getting stuck in local optima. More sophisticated optimization needs stronger mutations, i.e. higher mutation rates.

\subsection{Social Impact}

Our defined loss vector components have illustrative interpretations. For instance, $L_+$ minimizes food or crop waste by overshoot reduction, whereas $L_-$ minimizes logistic costs accompanied by greenhouse gas emissions. By construction, our hierarchy prefers food waste prevention to emission reduction. Firstly, it reduces the food waste by accepting additional emission. Secondly, the emission is reduced by distributing the yield preferably efficient. Given the hierarchy, no new food waste is created thereby.

Furthermore, the bootstrap shows that the optimization just for one GDU forecast \textit{overfits} on that single scenario. This is visible by the resulting error bars in Figure~\ref{fig:harvest_results}. Optimizing for many GDU forecast realizations is expensive due to many loss evaluations. Nevertheless, we observe that the uncertainty propagation from GDU forecast to harvest yield still stays within standard deviation. Thus, our optimization is robust against environmental forecast uncertainties and can schedule the harvest reliably. Additionally, the optimization can be adjusted on-the-fly, e.g. once a more precise forecast (harvest matrix) is available.

We introduce the overshoot (undershoot) reduction ratio $R_o$ ($R_u$) that describes, to which extent we can reduce food waste and emissions with respect to the unoptimized case given in the data. As shown in Figure~\ref{fig:harvest_results}, we can reduce \SIrange{62}{100}{\percent} of over- and undershoot for both sites and all scenarios which is a drastic improvement. Using our method, one could thus tackle global food waste on a supply chain as well as on a farming level, which additionally contributes to a more efficient agricultural land use.


\bibliography{references}
\clearpage

\appendix
\section{Appendix}

\begin{figure}[t]
	\centering
	    \includegraphics[width=\columnwidth]{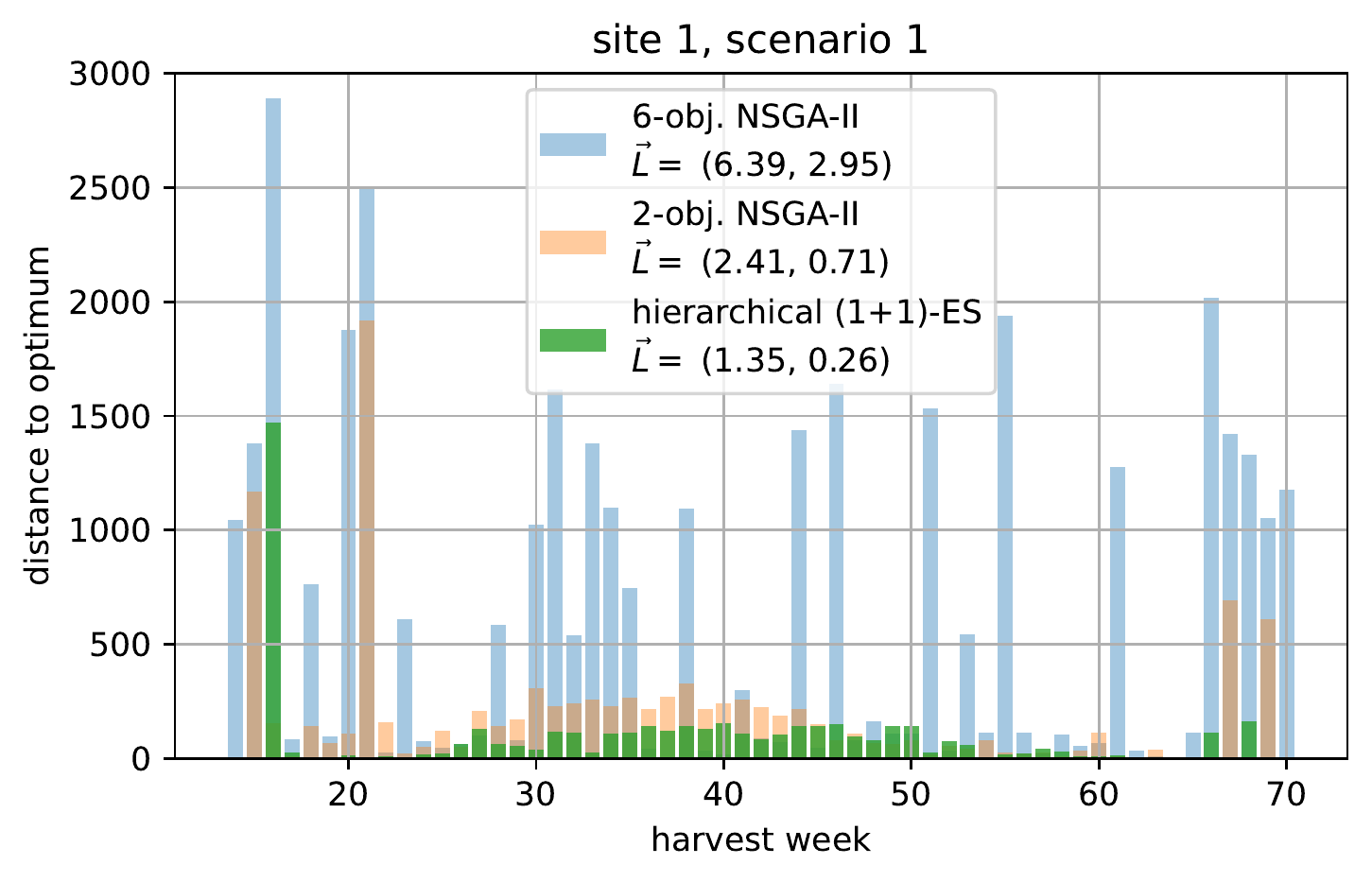}
	\vfill
		\includegraphics[width=\columnwidth]{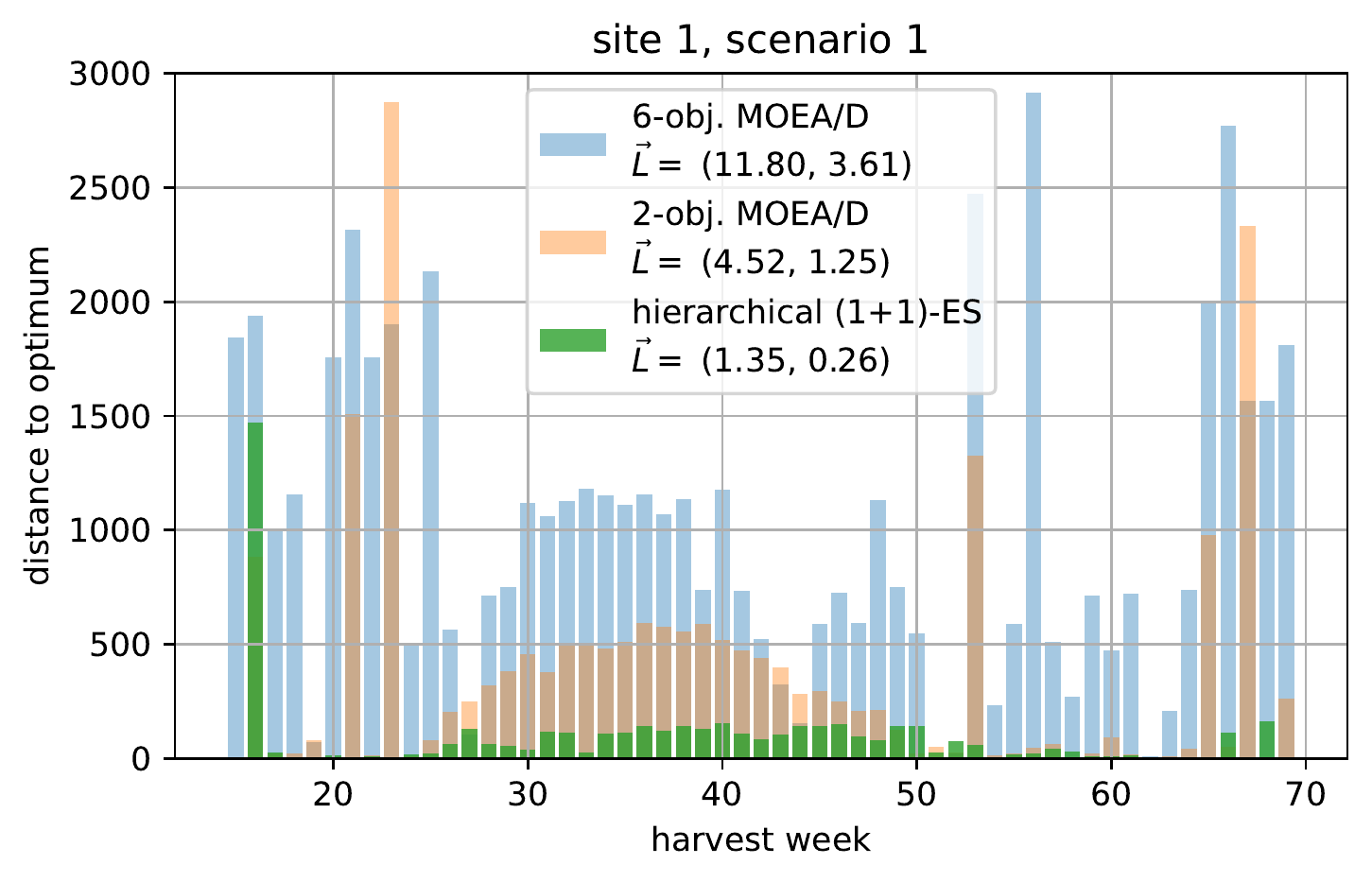}
	\vfill
		\includegraphics[width=\columnwidth]{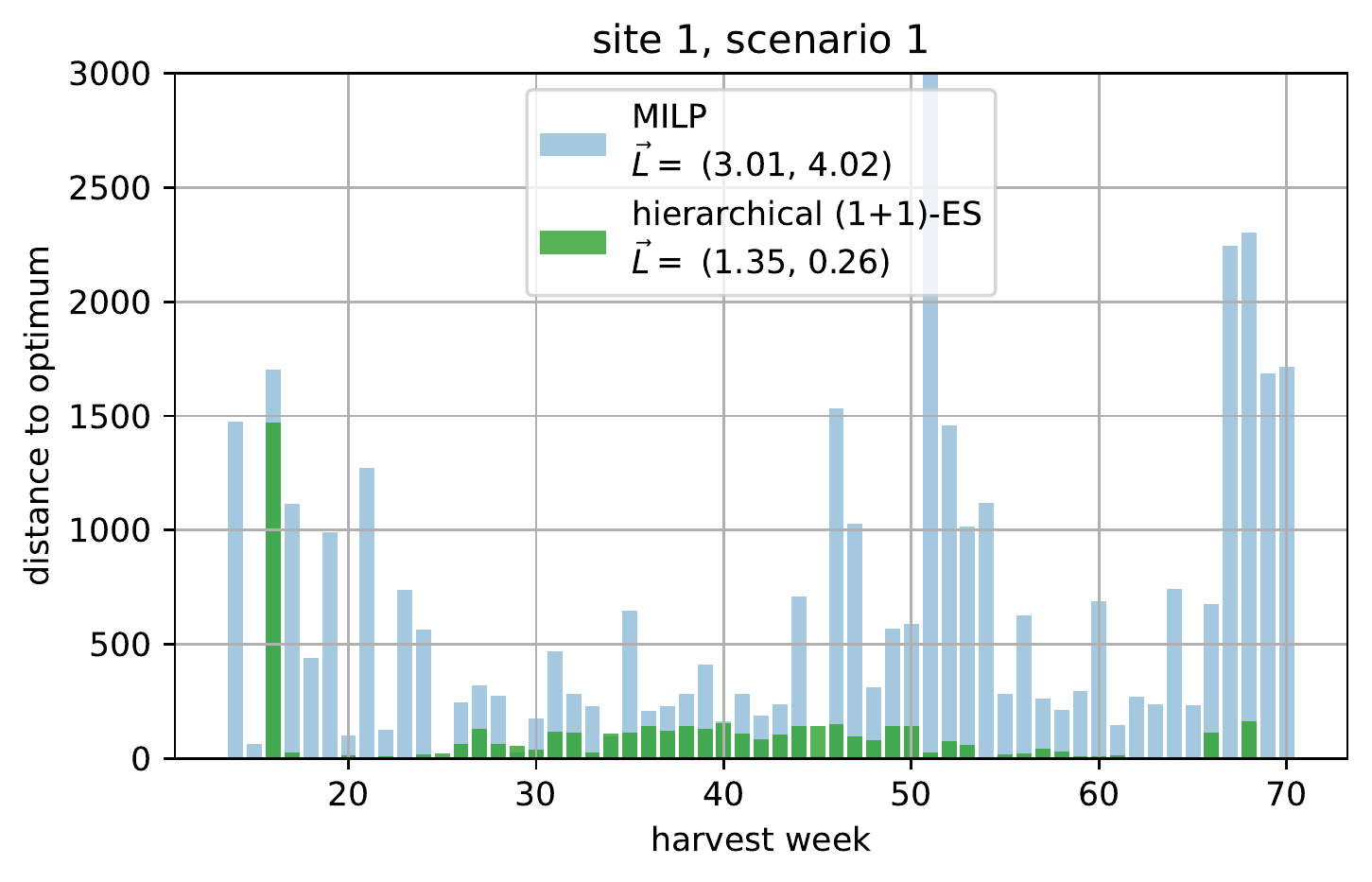}
	\caption{Comparison of different optimization strategies. For S1 and site 1, we compare our method with two established EAs, namely NSGA-II (top) and MOEA/D (middle). In addition, the bottom plot shows a comparison with an exact method represented by formulation as a mixed-integer linear program (MILP) solved with CPLEX\textregistered{}. Evolutionary approaches are stopped after $10^6$ evaluations. The bars show the distance to the next optimum, i.e. capacity limit or zero harvest. Thus, the lower the bars, the better the optimization state. For the evolutionary approaches, we show the 6-objective (blue bars) and the 2-objective (orange bars) optimization with the respective algorithm. Our $(1+1)$-ES method, shown by the green bars in all plots, has the lowest overall deviations and, thus, outperforms the reference methods. The hierarchical loss, given in the legend, is only used for the 2-objective and our approaches.}
	\label{fig:comp}
\end{figure}

\subsection{Comparison between Many-, Multi-, and Hierarchical Single-objective Optimization}\label{sec:baseline_comp_ea}
In order to show that transferring the harvest scheduling into a single-objective optimization problem improves the convergence, we test our $(1+1)$-ES against the commonly known algorithms NSGA-II and MOEA/D as a baseline. To get a proxy for the speed of convergence, we feature how the methods perform in a short-run optimization stopped after $10^6$ seen populations. We compare a many-objective approach with six efficiency and continuity criteria with our hierarchical single-objective approach. Additionally, a multi-objective approach by using the two components of the loss vector $\vec{L}$ as (coordinative) objectives is evaluated. The upper and middle plots of Figure~\ref{fig:comp} show the results for the three stated problem formulations. The goodness of optimization is shown by considering distance to the next optimum (capacity limit or zero harvest) for each harvest week. Thus, lower bars represent better optimization. As expected, our approach shows a faster convergence to a rather preferable interim solution. Certainly, the solutions can be further improved by a long-run optimization.

\subsection{Comparison with Exact Methods}\label{sec:baseline_comp_exact}
Indeed, it is possible to formulate our problem as a mixed-integer linear programming (MILP) problem. However, as stated in the Introduction chapter, exact methods like MILP need to have objective functions that are less generic than those of, for instance, (meta-)heuristic approaches like EAs. Thus, we have to deal with several limitations that can not be considered by the MILP method. For instance, the concave loss function cannot be minimized by the solver. Additionally, our scenarios with variable target capacity cannot be represented by an appropriate objective function either. The lower plot in Figure~\ref{fig:comp} shows the result, if we solve the S1 scenario (fixed target capacity) of our harvest scheduling problem by using the commercial solver CPLEX\textregistered{}. As an objective function, we are limited to use the summed squared error over all weeks between harvest yield and target capacity. As well as for the common two compared EA methods, our method clearly outperforms the MILP method. Note that the given MILP solution is already the best solution we can get at this point while our method was stopped in an early state. This vividly shows that complex problems like harvest scheduling can be solved better if the used optimization method allows for including sophisticated objectives and prior knowledge.

\subsection{Code Availability}
The dataset used in this work is restricted by copyright. Nevertheless, we generated a similar dataset with the same characteristics. Example code to reproduce optimization results and comparison with NSGA-II and MOEA/D is available at \url{https://github.com/mrcgndr/harvest_schedule_optimization}.

\end{document}